\documentclass[prb,twocolumn,showpacs]{revtex4}
\usepackage{graphicx,amsmath}

\newcommand{\figwidth}{0.95\columnwidth}
\newcommand{\eq}[1]{Eq.(\ref{#1})}
\newcommand{\fig}[1]{Fig.~\ref{#1}}
\newcommand{\avg}[1]{ {\langle #1 \rangle} }
\newcommand{\olcite}[1]{Ref.~\onlinecite{#1}}

\newcommand{\rhocr}{\rho_{\rm cr}}
\newcommand{\rhocra}{\rho_{\rm A,cr}}

\newcommand{\na}{N_{\rm A}}
\newcommand{\nb}{N_{\rm B}}
\newcommand{\za}{z_{\rm A}}
\newcommand{\zb}{z_{\rm B}}

\newcommand{\zbcr}{z_{\rm B,cr}}
\newcommand{\pc}{P_L(\na,\nb|\za,\zb)}

\newcommand{\rl}{\rho_{\rm L}}
\newcommand{\rv}{\rho_{\rm V}}

\newcommand{\ZBCR}{0.93791}
\newcommand{\RCR}{0.7486 \pm 0.0002}

\begin{document}

\title{Critical behavior of the Widom-Rowlinson mixture: coexistence
diameter and order parameter}

\author{R. L. C. Vink}
\affiliation{Institut f\"ur Theoretische Physik II, Heinrich Heine
Universit\"at D\"usseldorf, Universit\"atsstra{\ss}e 1, 40225
D\"usseldorf, Germany}

\date{\today}

\begin{abstract} 

The critical behavior of the Widom-Rowlinson mixture [J.~Chem.~Phys.~{\bf
52}, 1670 (1970)] is studied in $d=3$ dimensions by means of grand
canonical Monte Carlo simulations. The finite size scaling approach of
Kim, Fisher, and Luijten [Phys.~Rev.~Lett.~{\bf 91}, 065701 (2003)] is
used to extract the order parameter and the coexistence diameter. It is
demonstrated that the critical behavior of the diameter is dominated by a
singular term proportional to $t^{1-\alpha}$, with $t$ the relative
distance from the critical point, and $\alpha$ the critical exponent of
the specific heat. No sign of a term proportional to $t^{2\beta}$ could be
detected, with $\beta$ the critical exponent of the order parameter,
indicating that pressure-mixing in this model is small. The critical
density is measured to be $\rho \sigma^3 = \RCR$, with $\sigma$ the
particle diameter. The critical exponents $\alpha$ and $\beta$, as well as
the correlation length exponent $\nu$, are also measured and shown to
comply with $d=3$ Ising criticality.

\end{abstract}



\maketitle

\section{Introduction}

The Widom-Rowlinson (WR) mixture \cite{widom.rowlinson:1970} is a simple
model of a fluid exhibiting phase separation. The model consists of A and
B particles that interact via simple pair potentials: the AA and BB pair
interaction is ideal, while AB pairs interact via a hard-core potential of
diameter $\sigma$. Upon increasing density, the WR mixture phase separates
into an A-rich and B-rich phase. In $d=3$ dimensions, computer simulations
agree that the corresponding universality class is that of the
three-dimensional (3D) Ising model \cite{shew.yethiraj:1996,
johnson.gould.ea:1997, gozdz:2005, buhot:2005}. There is, however, some
variation in the reported estimates of the critical density.

In order to probe fluid criticality using computer simulation, high
quality data are required. The latter are typically generated using Monte
Carlo (MC) methods, and considerable effort has been devoted to develop
efficient MC schemes. In \olcite{johnson.gould.ea:1997}, for example, a MC
cluster move is described for the WR mixture that is (nearly) free of
critical slowing down. However, as pointed out in \olcite{buhot:2005},
this type of move cannot be used to obtain the coexistence curve, which
may be problematic if one is interested in measuring, say, the critical
exponent $\beta$ of the order parameter. In \olcite{buhot:2005},
therefore, a different cluster move is formulated, based on
\olcite{buhot.krauth:1998}, which not only gives access to the coexistence
curve, but is also rejection free. Recently, the latter approach was
generalized to continuous potentials \cite{liu.luijten:2004}.

However, in addition to efficient MC sampling, of at least equal importance (if
not more) is the finite size scaling (FSS) algorithm used to extrapolate the
simulation data to the thermodynamic limit. FSS is essential because the
correlation length diverges at the critical point, and thus the true
thermodynamic limit is never captured in a finite simulation box, no matter how
efficiently it is simulated. For fluids, recently proposed {\it unbiased} FSS
algorithms formulated in the grand canonical ensemble seem particularly powerful
\cite{orkoulas.fisher.ea:2001, kim.fisher.ea:2003, kim.fisher:2004, kim:2005}.
The latter algorithms are unbiased in the sense that no prior knowledge of the
universality class is required: the critical point of the transition, as well as
some of the critical exponents, are an output. These unbiased algorithms were
used, for example, to resolve the universality class of the hard-core
square-well (HCSW) fluid and the restricted primitive electrolyte, both of which
were shown to exhibit 3D Ising critical behavior \cite{luijten.fisher.ea:2002,
kim.fisher.ea:2003}.

Unfortunately, it is not obvious how the MC cluster moves for the WR mixture 
generalize to the grand canonical ensemble. Grand canonical cluster moves for 
mixtures seem less common, but some have been presented in the literature 
\cite{orkoulas.panagiotopoulos:1994, vink.horbach:2004*1}. Of these, the MC move of 
\olcite{vink.horbach:2004*1}, a generalization of its canonical variant 
\cite{biben.bladon.ea:1996}, is readily applicable to the WR mixture. By using the 
MC move of \olcite{vink.horbach:2004*1}, a FSS analysis of the WR mixture using the 
above mentioned unbiased algorithms thus becomes possible. This, consequently, is 
the aim of the present work. Of particular interest is the coexistence diameter, 
whose critical behavior is governed by a very weak singularity that is challenging 
to extract from simulation data. Note that, at the time of writing, the approach of 
\olcite{kim.fisher.ea:2003} seems to be the only FSS algorithm available to extract 
the coexistence diameter correctly from simulation data. A correct description of 
the latter is required in order to reliably estimate the critical density 
\cite{kim.fisher:2004*b}. Since the coexistence diameter of the WR model has not 
received much attention in previous simulations, the present grand canonical 
approach is certainly warranted.

\section{Simulation method}

In the grand canonical ensemble, the volume $V$ and the fugacity $\za$
($\zb$) of species A (B) are fixed, while the particle numbers $\na$ and
$\nb$ fluctuate. The thermal wavelength is set to unity such that the
fugacity $z_\alpha$ directly reflects the number density $N_\alpha / V$ a
pure phase of $\alpha$ particles would have (recall that such a phase is
simply an ideal gas). In what follows, all remaining length scales are
expressed in terms of the hard-core diameter $\sigma$. The crucial
quantity is the finite-size grand canonical distribution $\pc$, defined as
the probability of observing a system containing $\na$ particles of
species A and $\nb$ particles of species B, at fugacities $\za$ and $\zb$,
with $L$ the lateral dimension of the cubic simulation box (the use of
periodic boundary conditions is assumed).

The distribution is obtained numerically in a grand canonical MC
simulation via the insertion and removal of particles. To simulate
efficiently, the cluster move of \olcite{vink.horbach:2004*1} is used;
$\pc$ is then obtained simply by maintaining a histogram. To overcome the
free energy barrier separating the phases, a biased sampling scheme is
also implemented \cite{virnau.muller:2004}. Here, the simulation is
divided into distinct intervals (called windows) each spanning a single A
particle, while $\nb$ is allowed to fluctuate freely inside each window
(the choice for A or B is arbitrary). The windows are then sampled
separately and successively. CPU time is divided such that the number of
``sweeps'' per window is the same for all windows. In this work, we say
that a sweep has passed when a given population of particles has
completely been replaced or updated by new ones. This is in contrast to
the more common approach of keeping the number of {\it attempted} MC moves
per window fixed. The latter approach, however, is less appropriate for
grand canonical simulations since the acceptance rate is typically density
dependent. Per window, approximately 1800 sweeps are generated. To obtain
a single distribution, an investment of around 7 CPU hours for a small
system ($L=8$), and 270 hours for a large system ($L=13$) is required. In
order to perform the subsequent FSS analysis, $\pc$ is measured for system
sizes $L=8-13$ at fugacities ranging from close to the critical point to
well into the coexistence region. Estimates of properties at intermediate
fugacities are obtained using the multiple histogram method
\cite{ferrenberg.swendsen:1989}. 

\section{Results}

As mentioned before, the WR mixture exhibits phase separation into an
A-rich and B-rich phase. Note that the B-rich phase may equally well be
regarded as being poor in A species. In this sense, then, phase separation
is analogous to liquid-vapor coexistence: the A-rich phase being the
liquid, the A-poor phase being the vapor, and the fugacity of the B
particles being inverse temperature (again, the choice for A or B is
arbitrary). A natural definition of the order parameter is therefore
$\Delta \equiv (\rl-\rv)/2$, with $\rl$ the number density of A particles
in the A-rich phase, and $\rv$ the number density of A particles in the
A-poor phase. Close to the critical point, the order parameter is expected
to scale as $\Delta \propto t^\beta$, with $t = \zb/\zbcr - 1$ the
distance from the critical point, and $\zbcr$ the critical ``inverse
temperature''. Similarly, the coexistence diameter can be written as $D
\equiv (\rl+\rv)/2$. The critical behavior of the latter is given by
\cite{kim.fisher.ea:2003*b}
\begin{equation}\label{eq:coex}
 D = \rhocra \left( 1 + A_{2\beta} t^{2\beta} 
 + A_{1-\alpha} t^{1-\alpha} + A_1 t \right),
\end{equation}
with $\rhocra$ the number density of A particles at the critical point,
and non-universal amplitudes $A_i$. For the 3D Ising universality class,
appropriate exponent values are $\beta \approx 0.326$ and $\alpha \approx
0.109$ \cite{fisher.zinn:1998}.

\subsection{Order parameter}

\begin{figure}
\begin{center}
\includegraphics[clip=,width=\figwidth]{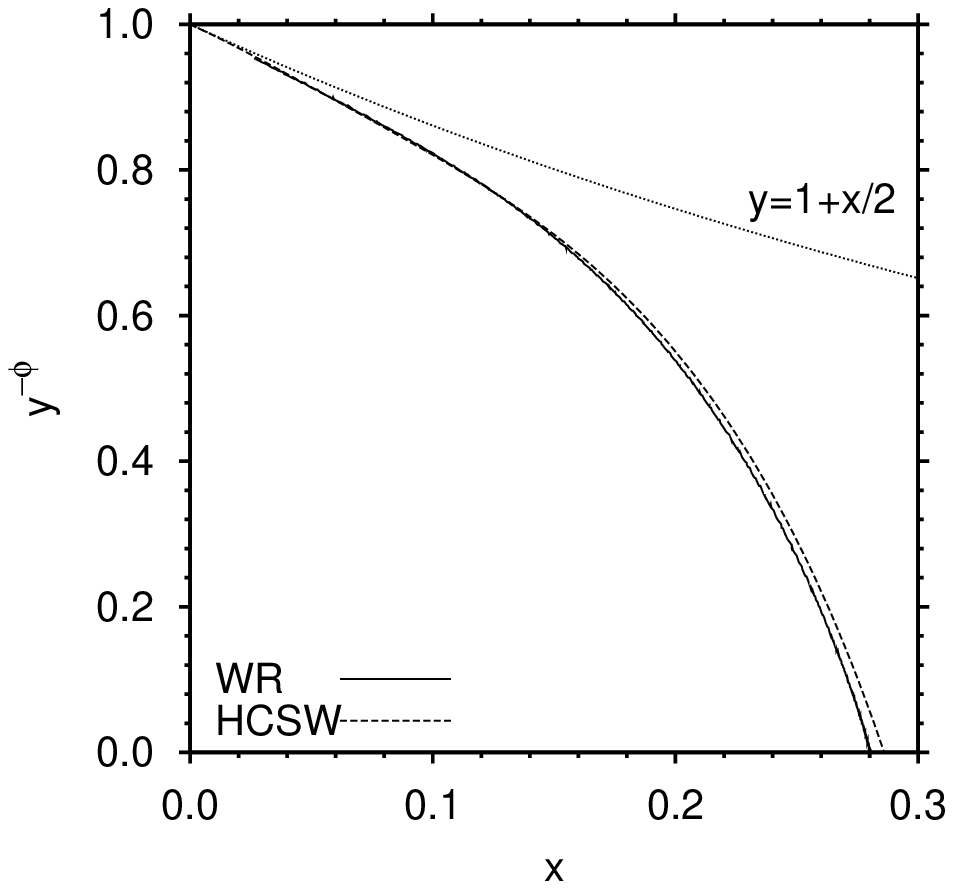}
\caption{\label{scaling}}
\end{center}
\end{figure}

\begin{figure}
\begin{center}
\includegraphics[clip=,width=\figwidth]{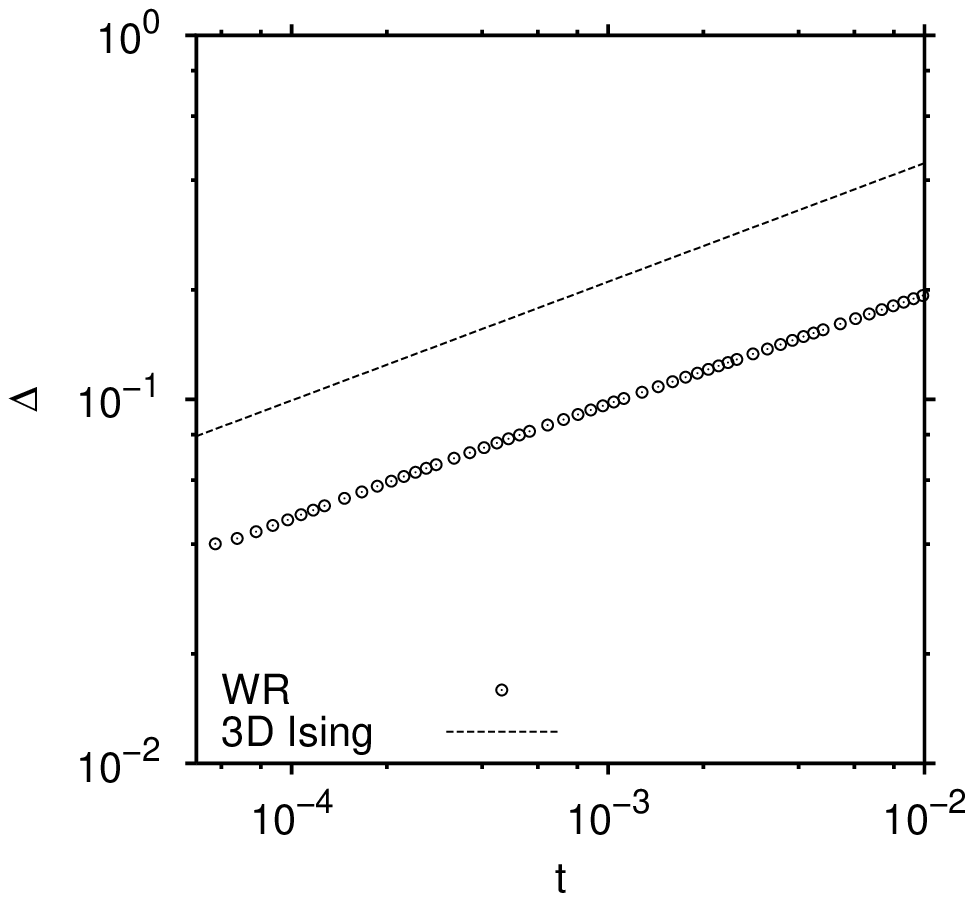}
\caption{\label{order}}
\end{center}
\end{figure}

To extract the order parameter, the FSS algorithm of \olcite{kim.fisher.ea:2003}
is used (for a more detailed description of the algorithm,
\olcite{kim.fisher:2004} is also highly recommended). The algorithm requires as
input the grand canonical distribution $\pc$ for at least three system sizes
$L$. Here, five system sizes $L=9,10,11,12,13$ are in fact used. Starting with
$\zb$ significantly above its critical value, the cumulant ratio $\avg{m^2}^2 /
\avg{m^4}$ is plotted as function of the average number density $\avg{\na}/V$,
with $m=\na-\avg{\na}$ (note that this plot is parameterized by the fugacity of
the other species $\za$). The resulting curve will reveal two minima, located at
$\rho^-$ and $\rho^+$, with respective values $Q^-$ and $Q^+$ at the minima.
Defining the quantities $Q_{\rm min} = (Q^+ + Q^-)/2$, $x = Q_{\rm min} \ln (4/e
Q_{\rm min} )$, and $y = (\rho^+ - \rho^-) / (2 \Delta)$, the points $(x,y)$
from the different system sizes should, in the limit far away from the critical
point, collapse onto the line $y=1+x/2$. Recall that $\Delta$ is the order
parameter in the thermodynamic limit at the considered fugacity $\zb$, precisely
the quantity of interest, which may thus be obtained by fitting until the best
collapse onto $1+x/2$ occurs. In the next step, $\zb$ is chosen closer to the
critical point, the points $(x,y)$ are calculated as before, but this time
$\Delta$ is chosen such that the new data set joins smoothly with the previous
one, yielding an estimate of the order parameter at the new fugacity. This
procedure is repeated all the way to the critical point, where $\Delta$
vanishes, leading to an estimate of the critical fugacity $\zbcr$. Moreover, the
procedure also yields $y$ as function of $x$. The latter scaling function is
universal within a universality class, and for the HCSW fluid can be found in
\olcite{kim.fisher.ea:2003}. Since the WR mixture belongs to the same
universality class, a similar curve should be found. The latter is verified in
\fig{scaling}, which shows $y$ as function of $x$ obtained in this work,
compared to the result of \olcite{kim.fisher.ea:2003}. The agreement is very
reasonable. From the vanishing of the scaling function, at $x_c = 0.280$, an
unbiased estimate of the critical fugacity $\zbcr = \ZBCR \pm 0.00004$ is
obtained. Note that $x_c$ is universal within a universality class. The estimate
reported here compares favorably to $x_c=0.286$ obtained for the HCSW fluid
\cite{kim.fisher.ea:2003}, and $x_c=0.296$ obtained for the 3D Ising model
\cite{tsypin.blote:2000}, providing additional confirmation that these systems
belong to the same universality class. Shown in \fig{order}, on double
logarithmic scales, is the order parameter $\Delta$ of the WR mixture as
function of the distance from the critical point $t$, where the above quoted
estimate of $\zbcr$ was used. The resolution of the present data is such that
$\Delta$ can be resolved down to $t \approx 5 \times 10^{-5}$. By fitting the
lowest few points in \fig{order} to $\Delta \propto t^\beta$, the critical
exponent is {\it measured} to be $\beta \approx 0.322 \pm 0.008$, which is
certainly compatible with the accepted 3D Ising value.

\subsection{Coexistence diameter}

\begin{figure}
\begin{center}
\includegraphics[clip=,width=\figwidth]{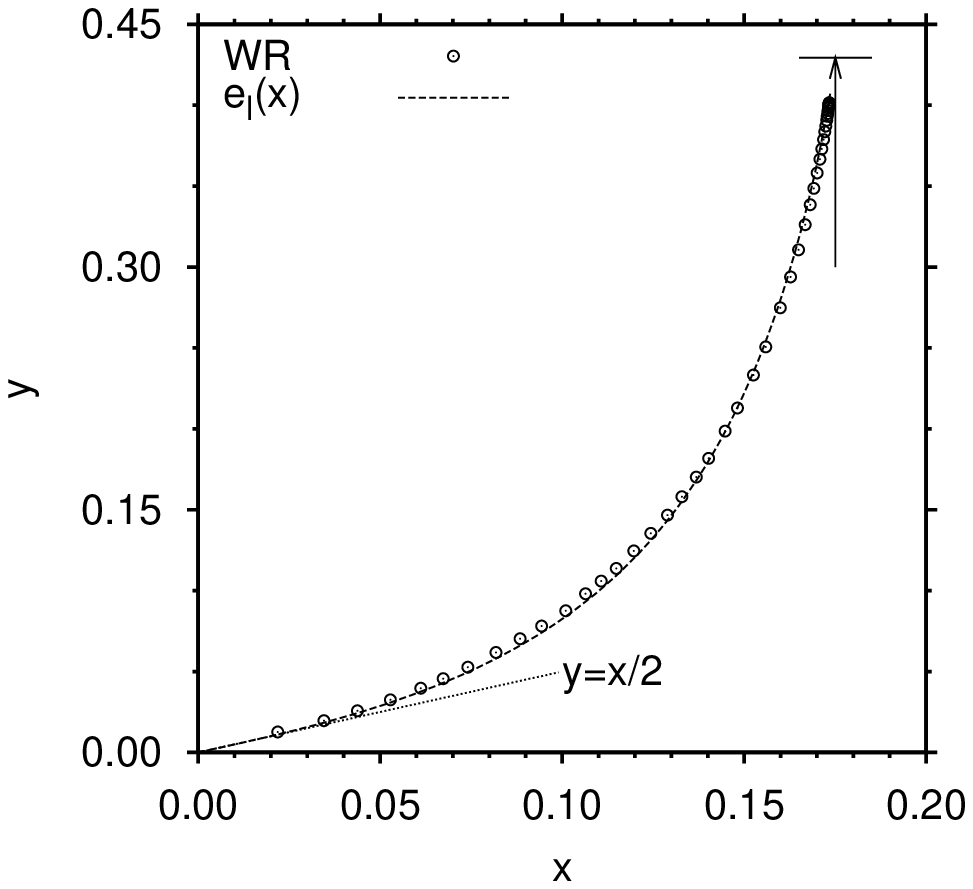}
\caption{\label{sc-diam}}
\end{center}
\end{figure}

To extract the coexistence diameter, the FSS algorithm of \olcite{kim:2005} is
used. The algorithm is similar in spirit to the previous one, in the sense that
it generates a scaling function $y=f(x)$, starting with data obtained well away
from the critical point, and then recursively working its way down toward
criticality. In \fig{sc-diam}, the scaling function of the diameter for the WR
mixture thus obtained is shown, where, as before, five system sizes $L=9-13$
were used. For $x \to 0$, this function is expected to approach $y=x/2$, which
indeed it does. In contrast to the order parameter, however, the scaling
function of the coexistence diameter is {\it not} universal \cite{kim:2005}.
Therefore, a direct comparison to scaling functions of other systems cannot, in
general, be carried out. Nevertheless, for systems with negligible pressure
mixing, such as the HCSW fluid and presumably also the WR mixture, the scaling
function is expected to be well described by the approximant \cite{kim:2005}
\begin{eqnarray}\label{eq:app}
 e_l(x) &=& C_l \left[ 1 - (1-\bar{x})^{1-\alpha} \right. \\
 && \times \left. \frac{ 1+s_1 \bar{x} + s_2 \bar{x}^2 + s_3 \bar{x}^3 }{
 1 + t_1 \bar{x} + t_2 \bar{x}^2 + t_3 \bar{x}^3} \right], \nonumber
\end{eqnarray}
with $\bar{x}=x/x_c$, $t_1 = s_1 - 1 + \alpha + x_c/2C_l$, and critical exponent
$\alpha \approx 0.109$. A fit to the WR data of \fig{sc-diam} shows that this is
indeed the case, with explicit parameter values $C_l = 0.429$, $x_c = 0.175$,
$s_1 = 4.50$, $s_2 = -5.72$, $s_3=0.12$, $t_1=3.81$, $t_2 = -9.08$ and
$t_3=4.25$. These values are remarkably consistent with estimates quoted in
\olcite{kim:2005} for the HCSW fluid. Note that $e_l(x)$ becomes singular
close to $x_c$, implied by the $(1 - \bar{x})^{1 - \alpha}$ factor in
\eq{eq:app}. The latter would yield a vertical tangent in the plot, at the arrow
in \fig{sc-diam}. The present simulation data, however, seem not to extend close
enough to the critical point to reach this regime.

\begin{figure}
\begin{center}
\includegraphics[clip=,width=\figwidth]{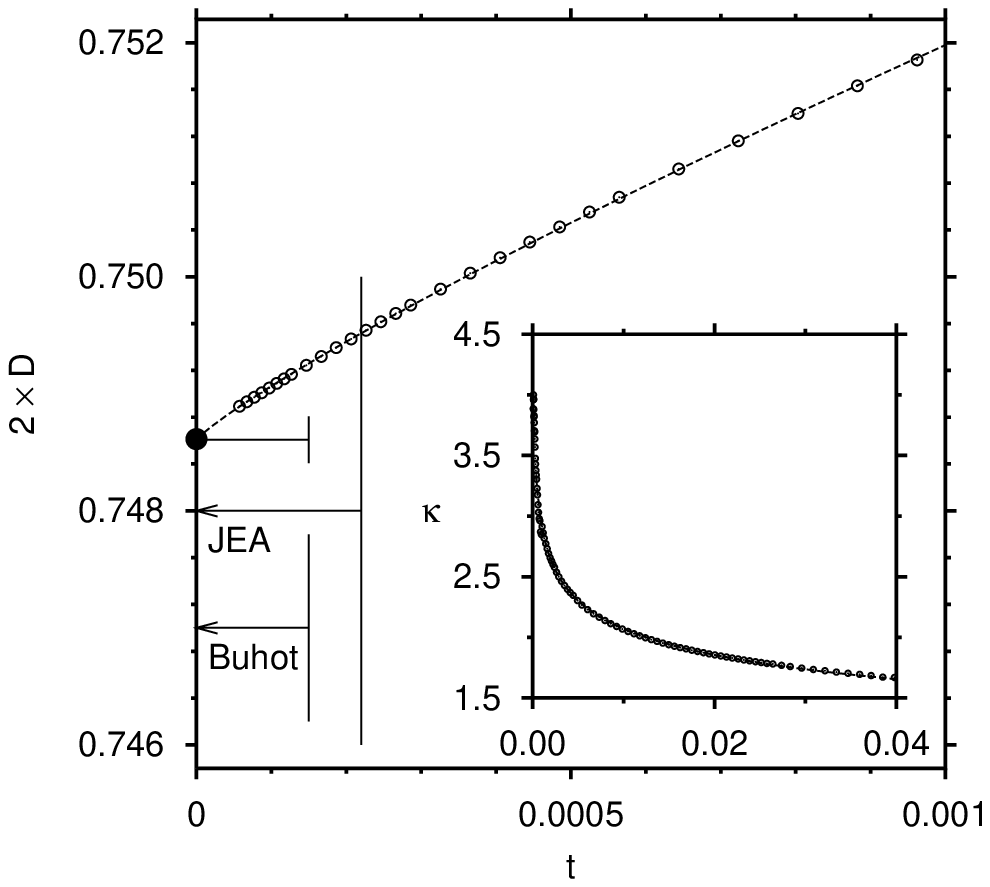}
\caption{\label{fit_diam}}
\end{center}
\end{figure}

The critical behavior of the coexistence diameter $D$ is shown in
\fig{fit_diam}, where $\zbcr = \ZBCR$ obtained in the previous paragraph
was used. In order to facilitate the comparison to other work, $2 \times
D$ is actually plotted. Symmetry considerations ensure equal numbers of A
and B particles at criticality, such that the {\it overall} critical
number density equals $\rhocr = 2 \rhocra$, which is the quantity usually
quoted in the literature. A fit to the asymptotic expansion of
\eq{eq:coex} yields $\rhocr = \RCR$, where the error reflects the
variation stemming from the range over which the fit is performed
(repeating the entire analysis leaving out the smallest system size yields
a similar result). The corresponding amplitudes read as $A_{2\beta}
\approx 0$, $A_{1-\alpha} = 2.76 \pm 0.07$ and $A_1 \approx -1.27 \pm
0.09$, implying that the singular behavior is dominated by $t^{1-\alpha}$.
This, in combination with the observation that the scaling function is
well described by $e_l(x)$, confirms that pressure mixing in the WR
mixture is small. The curvature of the diameter close to the critical
point thus reflects the $t^{1-\alpha}$ singularity. Since $1-\alpha$ is
close to unity, and the magnitudes of $A_{1-\alpha}$ and $A_1$ are
similar, the curvature is hard to see in \fig{fit_diam}. The singular
behavior of the diameter can be visualized nevertheless by plotting the
``inverse temperature'' derivative $\kappa = 2 \, d D/d t$ instead. In
case of singular behavior, $\kappa$ is expected to diverge when $t \to 0$,
see \eq{eq:coex}. Though not very precise, this procedure even allows for
an unbiased measurement of the exponent $\alpha$. The result is summarized
in the inset of \fig{fit_diam}, which shows $\kappa$ as function of $t$,
where again $\zbcr = \ZBCR$ in $t$ was used. The divergence is clearly
visible. By fitting the data to the form $\kappa = a_1 t^{-\alpha} + a_2$,
with fit parameters $a_i$ and $\alpha$, the specific heat exponent is
measured to be $\alpha = 0.11 \pm 0.02$, which is surprisingly close to
the 3D Ising value.

\subsection{Cumulant intersections}

\begin{figure}
\begin{center}
\includegraphics[clip=,width=\figwidth]{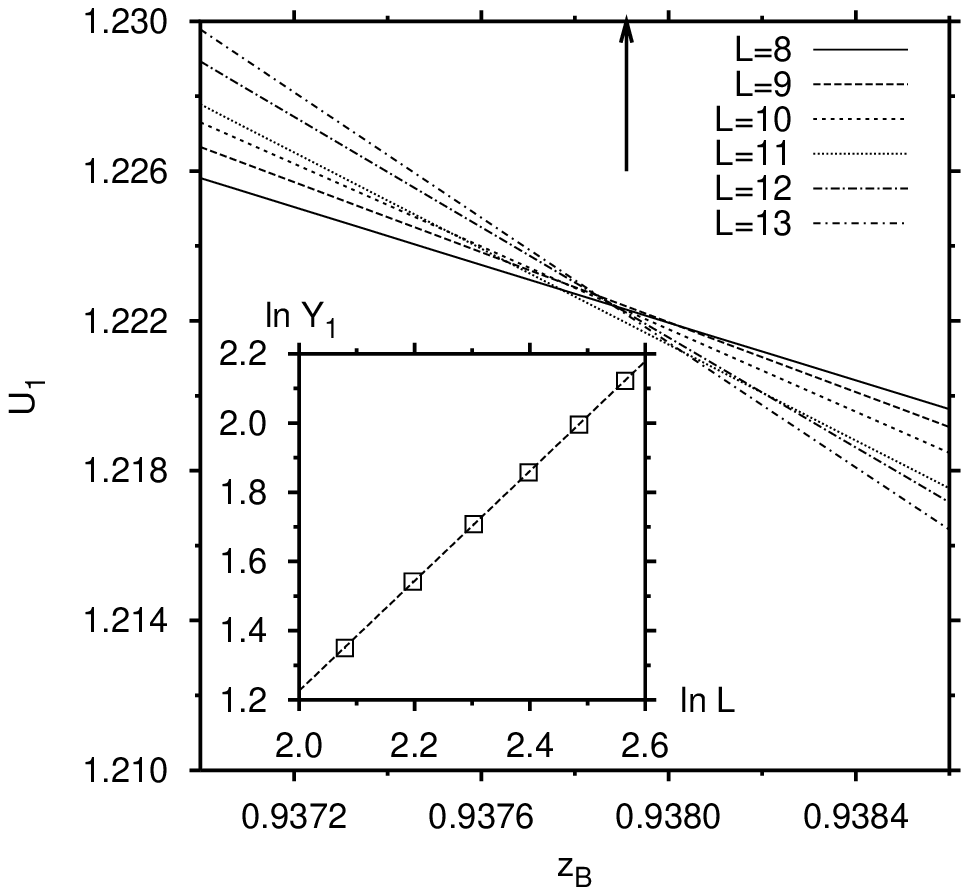}
\caption{\label{cumulant}}
\end{center}
\end{figure}

For the sake of completeness, and also to check the consistency of the results
obtained so far, the critical fugacity $\zbcr$ is measured again, but this time
around using the cumulant intersection approach \cite{binder:1981}. As was shown
by Binder \cite{binder:1981}, the (for example) first order cumulant $U_1 =
\avg{m^2} / \avg{|m|}^2$ becomes system-size independent at the critical point.
Plots of $U_1$ as function of $\zb$ for different system sizes are thus expected
to show a common intersection point, leading to an unbiased estimate of the
critical fugacity. Moreover, the cumulant value $Q_c$ at the intersection point
is universal, dependent only on the universality class. Shown in \fig{cumulant}
is the result of this procedure, on a rather fine scale. The resulting estimate
reads as $\zbcr = 0.9379 \pm 0.0004$, where the error reflects the scatter in
the intersection points. The latter is fully consistent with the previous, more
precise value, $\zbcr = \ZBCR \pm 0.00004$ (arrow in \fig{cumulant}). For the
critical value of the cumulant $Q_c \approx 1.223$ is obtained. This value
compares quite favorably to the estimate $Q_c = 1.2391 \pm 0.0014$ obtained in
large-scale simulations of the 3D Ising lattice model \cite{luijten:1999},
deviating from it by less than 2\%. The inset of \fig{cumulant} shows the slope
of the cumulant $Y_1 = | {\rm d} U_1 / {\rm d} \zb |$ at the critical fugacity,
as function of the system size $L$. It is expected that $Y_1 \propto L^{1/\nu}$
with $\nu$ the critical exponent of the correlation length. Although there is
some scatter in the intersection points, the cumulant slopes seem rather
constant over the range of \fig{cumulant}, and so it is expected that $\nu$ can
be obtained quite reliably nevertheless. Indeed, by performing a fit to the data
in the inset of \fig{cumulant}, the exponent is measured to be $\nu \approx
0.630 \pm 0.005$, in excellent agreement with the accepted 3D Ising value
$\nu_{\rm Is} \approx 0.630$ \cite{fisher.zinn:1998}.

\section{Discussion and Summary}

In this work, the critical behavior of the WR mixture was investigated
using grand canonical MC simulations and unbiased FSS algorithms. As
expected, the universality class of the transition is that of the 3D Ising
model. This was demonstrated by direct measurements of the exponents $\alpha$,
$\beta$ and $\nu$. Substantial indirect evidence has also been provided,
by comparing the scaling function of the order parameter and the
coexistence diameter to those of the HCSW fluid, as well as via the
universality of $Q_c$ at the cumulant intersection point.

The critical density obtained in this work $\RCR$ can be compared to other 
simulations. Shew and Yethiraj report $0.762 \pm 0.016$ \cite{shew.yethiraj:1996} 
using semigrand simulations \cite{deutsch.binder:1992, miguel.rio.ea:1995}, while 
Johnson {\it et al.}~(JEA) obtained $0.748 \pm 0.002$ \cite{johnson.gould.ea:1997}. 
More recent estimates are due to G\'o\'zd\'z, $0.759 \pm 0.019$ \cite{gozdz:2005} 
using the Bruce-Wilding field mixing technique \cite{bruce.wilding:1992}, and Buhot 
$0.7470 \pm 0.0008$ \cite{buhot:2005}. Of the above, JEA and Buhot are very close 
to the value reported here, see \fig{fit_diam}. The estimate of G\'o\'zd\'z, 
however, is higher. A possible explanation is the adapted FSS algorithm in the 
latter, which seems to overestimate the critical density in some cases 
\cite{kim.fisher:2004*b}. JEA also report an estimate of the critical fugacity 
$\zbcr \approx 0.9403$ for the largest system size considered by them 
\cite{johnson.gould.ea:1997}, but without a systematic FSS analysis of this 
quantity. This overestimates the present value significantly. Interestingly, these 
authors observe an {\it increase} of $\zbcr$ with system size, in disagreement with 
the present work.

Buhot, by using rejection-free cluster MC moves, is able to simulate
impressively large systems \cite{buhot:2005}, up to $L=100$, which exceeds
the typical system size of the present investigation by about one order of
magnitude. The improved accuracy of the critical density obtained in this
work may therefore seem surprising. It should be emphasized, however, that
critical phenomena are most conveniently studied in terms of a field
variable, such as temperature or, in the case of the WR mixture, the
fugacity. The ensemble used by Buhot, as well as the semigrand ensemble,
do not have access to the fugacity. Instead, in these ensembles, the
critical point is approached by varying the overall density
$\rho=(\na+\nb)/V$. This somewhat restricts the investigation of critical
phenomena because for every considered $\rho$, an explicit simulation
needs to be carried out. In grand canonical simulations, on the other
hand, one has access to the particle fugacities. This facilitates the {\it
extrapolation} of simulation data obtained at one set of fugacities to
different values via histogram reweighting
\cite{ferrenberg.swendsen:1989}. Clearly, the investigation of subtle
effects, such as the critical behavior of the coexistence diameter, is not
really feasible without such extrapolation methods. Note that the behavior
depicted in \fig{fit_diam} is not simply an ``artifact'' of the grand
canonical ensemble. In the semigrand ensemble, for example, the singular
behavior of the diameter leads to a renormalization of the critical
exponent $\beta$ \cite{fisher:1968}. Assuming negligible $A_{2\beta}$ in
\eq{eq:coex}, one obtains $\rho/\rhocr-1 \propto t^{1-\alpha}$ close to
the critical point. Combining this with the critical power law of the
order parameter $\Delta \propto t^\beta$ and eliminating $t$, yields
$\Delta \propto (\rho/\rhocr-1)^{\beta^\star}$, with renormalized exponent
$\beta^\star = \beta/(1-\alpha)$. The latter renormalized exponent has
been confirmed experimentally \cite{chen.payandeh:2000}, and should, in
principle, also show up in the WR mixture when, as mentioned above, the
critical point is approached by varying $\rho$.

Needless to say, the WR mixture has also been studied by theoretical
means, using for example density functional theory \cite{schmidt:2001},
and integral equations \cite{shew.yethiraj:1996, yethiraj.stell:2000}.
These investigations, however, all pertain to the mean-field level. As
such, quantitative agreement with computer simulations close to
criticality is not to be expected, and a comparison is consequently not
carried out.

\acknowledgments

This work was supported by the {\it Deutsche Forschungsgemeinschaft} under
the SFB-TR6 (project section D3). I also thank K.~Binder for an initial
reading of the manuscript.

\bibstyle{revtex}
\bibliography{mainz}


\newpage
\section{Figure captions}

FIG. 1: Scaling function of the order parameter. Following the convention of 
\olcite{kim.fisher.ea:2003}, the scaling function is raised to a negative 
exponent, with $\phi=1/\beta$ and $\beta=0.326$. The solid curve is the result 
obtained in this work for the WR mixture; the dashed curve is the HCSW result of 
\olcite{kim.fisher.ea:2003}. Also shown is the exact small $x$ limiting form $y 
= 1 + x/2$.

FIG. 2: Order parameter of the WR mixture as function of the
distance from the critical point. The dashed line has a slope
$\beta=0.326$, corresponding to the 3D Ising exponent.

FIG. 3: Scaling function of the coexistence diameter for
the WR mixture. Open circles show simulation results obtained using the
FSS algorithm of \olcite{kim:2005}. The dashed curve is a fit to the
simulation data using the approximant of \eq{eq:app}. Also shown is the 
exact small $x$ limiting form $y=x/2$.

FIG. 4: Coexistence diameter of the WR mixture as
function of the distance from the critical point. Open circles are
simulation results obtained in this work using the FSS algorithm of
\olcite{kim:2005}. The dashed curve is a fit to \eq{eq:coex}. The black
dot marks the critical density obtained from the fit, where the vertical
line indicates the uncertainty. The arrows mark estimates of $\rhocr$
reported in \olcite{johnson.gould.ea:1997} (JEA) and \olcite{buhot:2005}
(Buhot), where the vertical lines again indicate the uncertainty. The
inset shows $\kappa$ as function of $t$. Open circles are simulation
results; the dashed curve, which essentially overlaps the simulation data,
is a three-parameter fit of the form $\kappa = a_1 t^{-\alpha} + a_2$,
with fit parameters $a_i$ and $\alpha$.

FIG. 5:  Cumulant analysis of the WR mixture. Shown is
the first order cumulant $U_1$ as function of the fugacity $\zb$ for
various system sizes $L$ as indicated. The inset shows the slope of the
cumulant $Y_1$ at the critical point as function of $L$. All data were
obtained along the symmetry locus $\za=\zb$.


\newpage

\newpage

\newpage

\newpage

\newpage

\end{document}